\documentclass{ws-ijmpcs}

\newcommand{\dtres}{d^{\hspace{0.1mm} 3}\hspace{-0.5mm}}

\newcommand{\be}{\begin{eqnarray}}
\newcommand{\ee}{\end{eqnarray}}
\newcommand{\nn}{\nonumber}

\newcommand{\ppvec}{\vec{p}^{\;\prime}}

\begin{document}

\markboth{E. Oset et al.}
{The X(3872) and other X,Y,Z resonances as hidden charm meson-meson molecules}

%
\catchline{}{}{}{}{}
%

\title{\bf THE X(3872) AND OTHER X,Y,Z RESONANCES AS HIDDEN CHARM MESON-MESON MOLECULES}

\author{E. Oset$^{1,2}$, D. Gamermann$^{1,2}$, R. Molina$^{1,2}$, J.M. Nieves$^2$, E. Ruiz Arriola$^3$, T. Branz$^4$ and Wei Hong Liang$^5$}

\address{\it $^1$Departamento de Fisica Teorica,
Institutos de Investigacion de Paterna, Aptdo. 22085, 46071 Valencia, Spain\\
$^2$IFIC, Centro Mixto Universidad de Valencia-CSIC,
Institutos de Investigacion de Paterna, Aptdo. 22085, 46071 Valencia, Spain\\
$^3$Departamento de Fisica Atomica, Molecular y Nuclear, Universidad de Granada, E-18071 Granada, Spain\\
$^4$Institut fuer Theoretische Physik, Universitaet Tuebingen, Auf der Morgenstelle 14, D-72076 Tuebingen, Germany\\
$^5$Physics Department, Guangxi Normal University, Guilin 541004, China\\
}

\maketitle

\begin{history}
\received{Day Month Year}
\revised{Day Month Year}
\end{history}

\begin{abstract}
We report on some ideas concerning the nature of the X(3872) resonance and the need for approximately equal charged and neutral components of $D \bar{D}^* +cc$.
Then we discuss how some hidden charm states are obtained from the interaction between vector mesons with charm and can be associated to some of the charmonium-like X,Y,Z states. Finally we discuss how the nature of these states could be investigated through different types of radiative decay.

\keywords{Hidden charm molecules; X(3872); X,Y,Z.}
\end{abstract}

\ccode{PACS numbers: 14.40.Rt, 12.40.Vv, 13.75.Lb, 14.40.Lb}

\section{The X(3872) as a molecule of $D \bar{D}^* +cc$.}	

The fact that the mass of  the X(3872) resonance is so close to the $D^0 \bar{D}^{0*}$  has stimulated much work around the idea that it is a molecular state of hidden charm with these pseudoscalar-vector components \cite{Swanson:2003tb,Voloshin:2004mh,Braaten:2005ai,Dong:2009yp}. Although the charged components of these states could also play a role, the idea is that the neutral component is so little bound that it stretches far away and the probability to find this component is much larger than that of the charged components, which is confined because of the extra 7 MeV binding. The argument is intuitive but the short range nature of the strong interactions has as a consequence that only the wave function close to the origin is relevant for processes involving the strong interaction. Then the wave function around the origin is the relevant magnitude and not the absolute probability to find a certain component. Properties related to the isospin of the resonance will then be tied again to the wave functions at the origin. A theoretical study done in  \cite{Gamermann:2009fv} takes the model of \cite{Gamermann:2007fi}, where the X(3872) appears as a dynamically generated resonance in isospin I=0, and, by studying the system in charge basis, finds that the residues at the pole for the neutral and charged components are remarkably equal, thus indicating that one has an I=0 state. It is possible to address many observables in terms of these couplings and the equality of the couplings for the charged and neutral components would determine the behavior of the resonance as an object of isospin zero. In the same work it was discusses that with this solution it was possible to understand the ratio 

\be
\frac{{\cal B}(X\rightarrow J/\psi\pi^+\pi^-)}
{{\cal B}(X\rightarrow J/\psi\pi^+\pi^-\pi^0)}= 1.0\pm 0.4\pm 0.3\label{ratiopipi} \, , 
\ee
  Since the two pions in the numerator are known to be mostly a $\rho$ meson and the three pions in the denominator are mostly coming from an $\omega$ meson, it looks like the isospin zero option for the X indicates a strong isospin breaking in this reaction. Yet, it was shown in  \cite{Gamermann:2009fv} that this is not the case because the large width of the $\rho$ provides a large phase space for the reaction in the numerator compared to that in the denominator, which compensates for a small violation of isospin. Hence, a small isospin violation due to the different masses of the charmed mesons and vectors is sufficient, according to \cite{Gamermann:2009fv}, to produce a ratio for eq. (\ref{ratiopipi}) in agreement with experiment.  
   The study of the wave functions of the meson-vector components of the X(3872) resonance was done in \cite{Gamermann:2009uq}. For this we assumed a potential in momentum space of the type

\be
\langle \ppvec|V|\vec{p}\,\rangle =& V(\ppvec,\vec{p}\,)=& v \,\Theta(\Lambda-p)\Theta(\Lambda-p^\prime)\label{potms}
\ee
 which, via the Lippmann Schwinger equation, leads also to a factorized T-matrix 
\be
\langle \vec{p}\,|T|\ppvec\rangle
&=&\Theta(\Lambda-p)\Theta(\Lambda-p^\prime)\, t \label{eq:thetas}
\ee
 with $t$ given by $t=v+v\,G\,t$, or equivalently,

\be
  t = \frac{v}{1-vG} \label{tmat2}; \qquad {\rm with}~~
G&=&\int_{p<\Lambda}\dtres p\frac{1}{E-m_1-m_2-\frac{\vec{p}^{\,2}}{2\mu}} \label{eq13}
\ee

This equation is easily generalized to coupled channels and one finds the remarkable result that 
\be
g_i=\hat{\psi_i} / G^{\alpha}_{ii} \nn ; ~~ g_iG_{ii}^\alpha=\hat{\psi_i}
\label{eq75}
\ee
where $\alpha$ stands for the bound state of the system, and  $g_i$ is the coupling of the state to the channel $i$, determined from the residues of the scattering amplitudes in the pole of the bound state, and with interesting relationships with the derivatives of $G_{ii}$ \cite{Gamermann:2009uq}. The function  $G_{ii}$ is the loop function which appears in the scattering equation, eq. (\ref{eq13}), and $\hat{\psi_i}$ is defined as 
\be
(2\pi)^{3/2}\psi_i(\vec{0}\,)=&\hat{\psi_i} \label{eq53a} 
\ee
This result shows the relationship of the couplings to the wave function at the origin and one can see that either the couplings or the wave functions around the origin (see \cite{Gamermann:2009uq} for the case where there is smooth function substituting the sharp $\theta$ function, where the wave function at the origin is substituted by an average of the wave function at short distances).
   The findings of \cite{Gamermann:2009fv,Gamermann:2009uq} have been very useful to understand the meaning of the couplings and their role in strong interaction processes, plus their relationship to the wave functions and the relationship of the isospin of a multichannel state to the wave functions of the components close to the origin, rather than to their probability. They were also useful to understand that the role of the charged components in the X(3872) is essential and the picture with only neutral components does not stand up for a fair comparison with experiment. Indeed, the ratio of eq. (\ref{ratiopipi}) becomes about 50 times larger than experiment if only the neutral components are chosen.

\section{A description of some X,Y,Z resonances as bound states of hidden charm vector-vector states.}
In \cite{Molina:2009ct} the  $D^* \bar{D}^*$ system was studied and the scattering amplitudes produced poles which could be associated to hidden charm meson states. These meson states could be associated in some cases to recently claimed charmonium-like X,Y,Z states which, however, do not stand a neat association to expected standard charmonium states. These states obtained were a generalization of the case of the interaction of light vector-light vector \cite{Molina:2008jw,Geng:2008gx} or a charmed vector with a light one,
without strangeness \cite{Molina:2009eb} or with strangeness \cite{Molina:2010tx}. In all the previous cases one finds poles in the scattering amplitudes which can be associated to states already discovered, with predictions in a few cases for states not yet observed. The success of these cases, where predictions for the resonances properties could be done and contrasted favorable with experiment\cite{Oset:2010zzb}, gives us a certain confidence that the states predicted can indeed be associated to some of the discovered ones. In the Table that follows, which we borrow from \cite{Olsenmult}, we list some of these resonances and their properties.

\begin{table}[ht]
\begin{center}
\begin{tabular}{c|c|c|c|c|c}
State & M (MeV) & $\Gamma$ (MeV) & $J^{PC}$ & Decay modes & Production modes \\
\hline
\hline
$Z(3940)$& $3929\pm5$ & $29\pm 10$ & $2^{++}$ & $D\bar{D}$ & $\gamma\gamma$ \\
\hline
$X(3940)$ & $3942\pm 9$ &$37\pm 17$ & $J^{P+}$ & $D\bar{D}^*$ & $e^+ e^-\to J/\psi X(3940)$\\
\hline
$Y(3940)$ & $3943\pm 17$  & $87\pm 34$& $J^{P+}$ & $\omega J/\psi$ & $B\to KY(3940)$\\
 & $3914.3^{+4.1}_{-3.8}$& $33^{+12}_{-8}$& & & \\
\hline
$X(4160)$ & $4156\pm 29$ & $139^{+113}_{-65}$& $J^{P+}$ & $D^{*}\bar{D}^*$&$e^+e^-\to J/\psi X(4160)$\\
\hline
\end{tabular}
\end{center}
\caption{Properties of the candidate XYZ mesons.}
\label{tab:new}
\end{table}

On the other hand, the theoretical study of \cite{Molina:2009ct} finds poles at different positions and with certain quantum numbers of spin and isospin. The properties of these states, as well as their coupling to the different building channels are given in the following Tables.

\begin{table}[ht]
 \begin{center}
\centerline{$\sqrt{s}_{pole}=3943 + i 7.4$, $I^G[J^{PC}]=0^+[0^{++}]$}
\vspace{0.5cm}
\begin{tabular}{ccccc}
\hline
$D^*\bar{D}^*$&$D^*_s\bar{D}_s^*$&$K^*\bar{K}^*$&$\rho\rho$&$\omega\omega$\\
\hline
\hline
$18810-i 682 $&$8426+i 1933 $&$10- i 11$&$-22 + i 47$&$1348+ i 234 $\\
\hline
\end{tabular}\\
\vspace{0.5cm}
\begin{tabular}{ccccc}
\hline
$\phi\phi$&$J/\psi J/\psi$&$\omega J/\psi$&$\phi J/\psi$&$\omega\phi$\\
\hline
\hline
$-1000 -i 150$&$417+ i 64$&$-1429 - i 216$&$889+ i 196 $&$-215 - i107$\\
\hline
\end{tabular} 
\end{center}
\caption{Couplings $g_{i}$ in units of MeV for $I=0$, $J=0$.}
\label{tab:res1}
\end{table} 
\begin{table}[ht]
\begin{center}
\centerline{$\sqrt{s}_{pole}=3945 +i 0$, $I^G[J^{PC}]=0^-[1^{+-}]$}
\vspace{0.4cm}
\begin{tabular}{cccccccccc}
\hline
$D^*\bar{D}^*$&$D^*_s\bar{D}_s^*$&$K^*\bar{K}^*$&$\rho\rho$&$\omega\omega$&$\phi\phi$&$J/\psi J/\psi$&$\omega J/\psi$&$\phi J/\psi$&$\omega\phi$\\
\hline
\hline
$18489- i0.78 $&$8763+ i2 $&$11-i38 $&$0$&$0$&$0$&$0$&$0$&$0$&$0$\\
\hline
\end{tabular}
\end{center}
\caption{Couplings $g_{i}$ in units of MeV for $I=0$, $J=1$.}
\label{tab:res2}
\end{table} 
\begin{table}[ht]
\begin{center}
\centerline{$\sqrt{s}_{pole}=3922+i 26$, $I^G[J^{PC}]=0^+[2^{++}]$}
\vspace{0.5cm}
\begin{tabular}{ccccc}
\hline
$D^*\bar{D}^*$&$D^*_s\bar{D}_s^*$&$K^*\bar{K}^*$&$\rho\rho$&$\omega\omega$\\
\hline
\hline
$21100- i 1802 $&$1633+i 6797 $&$42+ i 14 $&$-75 +i 37$&$1558 + i 1821$\\
\hline
\end{tabular}\\
\vspace{0.5cm}
\begin{tabular}{ccccc}
\hline
$\phi\phi$&$J/\psi J/\psi$&$\omega J/\psi$&$\phi J/\psi$&$\omega\phi$\\
\hline
\hline
$-904 - i1783 $&$1783 +i 197$&$-2558 - i2289$&$918+ i2921 $&$91 -i 784$\\
\hline
\end{tabular}
\end{center}
\caption{Couplings $g_{i}$ in units of MeV for $I=0$, $J=2$.}
\label{tab:res3}
\end{table} 
\begin{table}[ht]
\begin{center}
\centerline{$\sqrt{s}_{pole}=4169+i 66$, $I^G[J^{PC}]=0^+[2^{++}]$}
\vspace{0.5cm}
\begin{tabular}{ccccc}
\hline
$D^*\bar{D}^*$&$D^*_s\bar{D}_s^*$&$K^*\bar{K}^*$&$\rho\rho$&$\omega\omega$\\
\hline
\hline
$1225- i 490 $&$18927- i 5524$&$-82 + i 30 $&$70+ i20$&$3 -i 2441$\\
\hline
\end{tabular}\\
\vspace{0.5cm}
\begin{tabular}{ccccc}
\hline
$\phi\phi$&$J/\psi J/\psi$&$\omega J/\psi$&$\phi J/\psi$&$\omega\phi$\\
\hline
\hline
$1257+ i 2866 $&$2681+ i 940$&$-866 + i 2752 $&$-2617 - i5151 $&$1012+ i 1522$\\
\hline
\end{tabular}
\end{center}
\caption{Couplings $g_{i}$ in units of MeV for $I=0$, $J=2$ (second pole).}
\label{tab:res4}
\end{table} 
\begin{table}[ht]
\begin{center}
\centerline{$\sqrt{s}_{pole}=3919+ i74$, $I^G[J^{PC}]=1^-[2^{++}]$}
\vspace{0.4cm}
\begin{tabular}{cccccc}
\hline
$D^*\bar{D}^*$&$K^*\bar{K}^*$&$\rho\rho$&$\rho\omega$&$\rho J/\psi$&$\rho\phi$\\
\hline
\hline
$20267-i 4975 $&$148- i33 $&$0$&$-1150 -i 3470 $&$2105+ i5978$&$-1067 -i 2514$\\
\hline
\end{tabular}
\end{center}
\caption{Couplings $g_{i}$ in units of MeV for $I=1$, $J=2$.}
\label{tab:res5}
\end{table}

  By looking for poles in the second Riemann sheet, we have found five resonances, three of which can be associated with the experimental data: The state $(3943,0^+[0^{++}])$ to the $Y(3940)$, the $(3922,0^+[2^{++}])$ to the $Z(3930)$ and the $(4157,0^+[2^{++}])$ to the $X(4160)$. There is no experimental counterpart for our state $(3945,0^-[1^{+-}])$, which is thus a prediction of our model.  These three states with mass around $3940$ MeV are basically composed by $D^*\bar{D}^*$, and  decay into pairs of light vectors like $K^*\bar{K}^*$, or light vector - heavy vector as $\omega J/\psi$, or $D \bar{D}$.
Our model predicts another state around $4160$ MeV, $(4157,0^+[2^{++}])$, which we identify with the $X(4160)$ state in base to the proximity of mass and widths and C-parity. This resonance has $J^{PC}=2^{++}$ and is mostly $D^*_s\bar{D}^*_s$. In the $I=1$ sector, the attraction is weak and we find only one resonance in the case of $J=2$, $(3912,1^-[2^{++}])$ which does not fit with any of the resonances known. 

   It is interesting to make predictions for these states before measurements are done. In this sense in \cite{Liang:2009sp} predictions are made for 
  radiative decay into $D^*$ and $\bar{D} \gamma$, or $D^*_s$ and $\bar{D}_s \gamma$, where it is shown that the nature of these resonances as vector vector molecules gives rise to a particular shape, something already noted before in  
\cite{Liu:2008tn}. Similarly, rates of radiative decay of these resonances into a $\gamma \gamma $ and $V \gamma$ are evaluated in \cite{Branz:2010rj}. There are no much data to compare, but the scarce existing data \cite{ueharaprl} indicate that 
the experimental X(3915) resonance fits well with the predicted resonance $(2^+, 3922)$, considered as the Z(3930) in \cite{Molina:2009ct}.


\end{document}